\documentclass{statsoc}

\usepackage[a4paper]{geometry}
\usepackage{comment}
\usepackage[textwidth=8em,textsize=small]{todonotes}
\usepackage{amsmath}
\usepackage{enumitem} %
\usepackage{natbib}
\usepackage{graphicx} %
\usepackage{xcolor}
\usepackage{amsfonts}
\usepackage{pdfpages}
\usepackage{hyperref}
\usepackage{bm} %

\usepackage{multicol}
\usepackage{multirow, rotating}

\usepackage{adjustbox}

\newcounter{subfigure}
\counterwithin{subfigure}{figure}

\graphicspath{{figures/}}

\usepackage{algorithm,algpseudocode,amsmath}

\makeatletter
\newcounter{phase}[algorithm]
\newlength{\phaserulewidth}
\newcommand{\setphaserulewidth}{\setlength{\phaserulewidth}}

\makeatother

\setphaserulewidth{.7pt}

\title[Shared Component Point Process Model]{A Shared Component Point Process Model for Urban\\ Policing}
\author{Claire Kelling} 
\address{Carleton College,
Northfield, MN,
United States.}
\email{ckelling@carleton.edu \\
  Center for Math \& Computing 225 \\ Northfield, MN 55057}
\author[C. Kelling \& M. Haran]{Murali Haran}
\address{Pennsylvania State University,
State College, PA,
United States.}

\begin{document}
\begin{abstract}

Newly available point-level datasets allow us to relate police use of force to other events describing police behavior. Current methods for relating two point processes typically rely on the spatial aggregation of one of the two point processes. We investigate new methods that build upon shared component models and case-control methods to retain the point-level nature of both point processes while characterizing the relationship between them. We find that the shared component approach is particularly useful in flexibly relating two point processes, and we illustrate this flexibility in simulated examples and an application to Chicago policing data.

\vspace{4pt}

\end{abstract}

\section{Introduction}

Event-level data with precise location information on phenomena such as police use of force incidents, police stops, and violent crime are increasingly available in many urban areas. Determining the causes and consequences of police use of force presents a growing challenge to criminologists and public policymakers
 as excessive use of force by police persists over time. We consider rich datasets on policing from Chicago. Existing work often incorporates high-level spatial information to study police behavior, such as indicator variables for the district where events occurs, rather than information about the exact locations \citep[cf.][]{antonovics2009new}. Point process methods allow us to incorporate detailed information about the precise location of events. Additionally, integrating related phenomena can also be helpful in studying police use of force. For example, the frequency and spatial distribution of both violent crimes and police stops may help characterize the spatial distribution of and factors influencing police use of force incidents. Police stops can give some preliminary information on where police are patrolling and the prevalence of violent crime is part of the characterization of the communities where police are patrolling. 

 Many methods exist for analyzing the relationship between different point processes, ranging from descriptive statistics to parametric and nonparametric methods. For instance, case-control models are often used to analyze the relationship between two types of events without aggregating either type of point. Specifically, the intensity of one point process, the ``control'' process, is used to scale the intensity of a second point process, the ``case'' process. In this paper we develop a new shared component model for point processes that provides a rich framework for analyzing the spatial relationship between two point patterns and avoids specifying case and control processes. We compare our shared component approach to existing models for relating two point processes, particularly case-control models. 

\noindent We summarize the main contributions of this paper below. %
\begin{itemize}[nosep]
    \item We propose a shared component model for two point processes %
    that allows for a spatial pattern that is unique to each point process as well as a pattern that is shared between the two point processes. This model builds upon the shared component model developed for areal data \citep{knorr2001shared}. We find through  simulated examples and application to Chicago policing data that our model is flexible and easy to interpret. 
    \item We study the use of a case-control model for point processes through simulation studies and applications to Chicago policing data. We consider two methods of estimating the intensity functions and regression coefficients in this context: logistic regression \citep[cf.][]{diggle2007second} and Bayesian estimation of a spatial intensity function. We find that care must be taken in choosing the estimation procedure and corresponding interpretation for this class of models. 
    \item We compare our shared component model to the case-control model and we find that the shared component model is computationally more complex but allows for additional flexibility and spatial structure when modeling the relationship between two point processes.
    \item Although our results are preliminary, we find that the shared component methodology allows us to effectively study the relationship between police use of force and police stops in Chicago. We illustrate a spatial pattern that is common to both point processes, south of downtown Chicago, and unique factors that impact the processes individually. Use of force events have a higher spatial intensity in the northern side of Chicago and police stops have a higher intensity in the southern side of the city, after accounting for a shared spatial pattern.
\end{itemize}

The paper is organized as follows. We describe areas of active research relating police use of force to other event datasets, such as police stops and violent crime, and the kinds of spatial analysis used in Section \ref{section_background}. In Section \ref{section_methods}, we describe statistical methods for relating two point processes. We give background information on case-control methods for point processes and describe our new shared component model for point processes. In Section \ref{section_data} we provide details about our Chicago data on police stops and police use of force used in this paper and also describe our simulated examples.  We apply the case-control and shared component models to Chicago policing data and simulated examples in Section \ref{section_results} and conclude with a discussion in Section \ref{section_discussion}.

\section{Background}
\label{section_background}

We begin by summarizing the kinds of policing data motivating this study along with related research questions in criminology. We also describe existing methodology that has been developed to analyze these datasets. 
 Examples of research questions include the following: What are possible factors that influence police use of force and how can additional rich point-level datasets help us better understand the spatial distribution of police use of force and police behavior more broadly? In the first subsection, we describe research relating police use of force to police stop data. In the second subsection, we similarly describe studies relating police use of force to another point-level dataset: violent crime. Finally, we illustrate current approaches to spatial analysis of these datasets. Most of these studies rely on spatial aggregation, though there has been some recent work on methods that reduce the amount of aggregation. 

\subsection{Police Use of Force and Police Stops}

Analysis of the spatial distribution of police stops can help indicate the baseline expectation for police use of force incidents. Although police stops do not represent all encounters with police, we can use police stops to give additional information about where use of force incidents are occurring at different rates from all police stops. There may be more use of force incidents in a given neighborhood not because of any characteristics of the neighborhood but rather because more police are patrolling the area.  \cite{ba2021role} show that when aggregating to large units over space and time, ``observed behavioral differences may simply reflect differing patrol environments, rather than differences in policing approaches.'' Proxies for police presence, such as the stops made by police, can be a useful tool in determining where police are patrolling, and can therefore help elucidate police use of force that is beyond where expected by an increased police presence. \cite{weisburst2019police} study use of force incidents as extensions of arrests and compare demographic patterns in both arrests and subsequent uses of force in the same incident. Police stops allow us to create a closer proxy to police presence than arrests, as not all stops involve arrests.

We note that police stops do not give us complete information on all individuals that police observe, as not all police encounters involve stops. The decision of officers to make a stop, out of all individuals that they encounter during a given patrol, may be biased. Studies have shown that it is important to consider bias in police stops when analyzing bias in police use of force \cite[cf.][]{knox2020administrative}. Failure to account for uneventful shifts may lead to inaccurate inferences on biases in policing \citep{knox2021revealing}. However, for the purposes of this study, we are analyzing only the spatial distribution between police stops and police use of force, not bias in outcomes of force. This framework could be expanded in the future to incorporate information about officers and civilians into the spatial model, for example through the two-stage framework developed by \cite{kelling2021two}.

\subsection{Police Use of Force and Violent Crime}

Data on violent crime is a second useful tool in determining the potential causes and consequences of police use of force. The amount of violent crime in a given area creates a neighborhood context that may have an important impact on police use of force, as officers may bring prior knowledge about the neighborhood. Through a vignette approach and survey responses, \cite{phillips2011police} finds that in an area with more violent crimes and higher crime rates as a percentage of the population, officers are more likely to perceive the use of unnecessary force as acceptable than in other areas, while controlling for other factors. In the criminology literature, there are hypotheses that places that have more violent crime and are disadvantaged are more likely to have more incidents of police use of force \citep{lawton2007levels, terrill2003neighborhood}. %
Specifically, there are many existing studies that find a weak positive relationship between violent crime and police use of force using the spatial distribution of violent crime \citep{lawton2007levels, lee2010examination, terrill2003neighborhood, lee2014impact}. 
 Yet, there is a conflicting hypothesis that prevalence of violent crime does not lead to an increase in police use of force \citep{slovak1988styles}. Other recent work shows that it may be ``impossible'' to determine whether violent crime in neighborhoods is a result or cause of policing in those neighborhoods \citep{simckes2021adverse}. We develop methods that allow us to analyze the relationship between different point processes, such as violent crime, police stops, and use of force, while preserving as much spatial and event-level information as possible and without specifying a causal direction that is implied by a case-control setup.

There are many possibilities for the ``baseline process'' to compare to use of force incidents, including violent crime, arrests, and stops, all of which provide slightly different information. For this study, we focus on the relationship between police stops and police use of force but this could be expanded in the future. We develop a method to study the impact of policing datasets on police use of force, while preserving the point-level nature of both datasets. %

\subsection{Use of Spatial Aggregation}

\begin{figure}[ht]
    \centering
    \includegraphics[width = 0.75\textwidth]{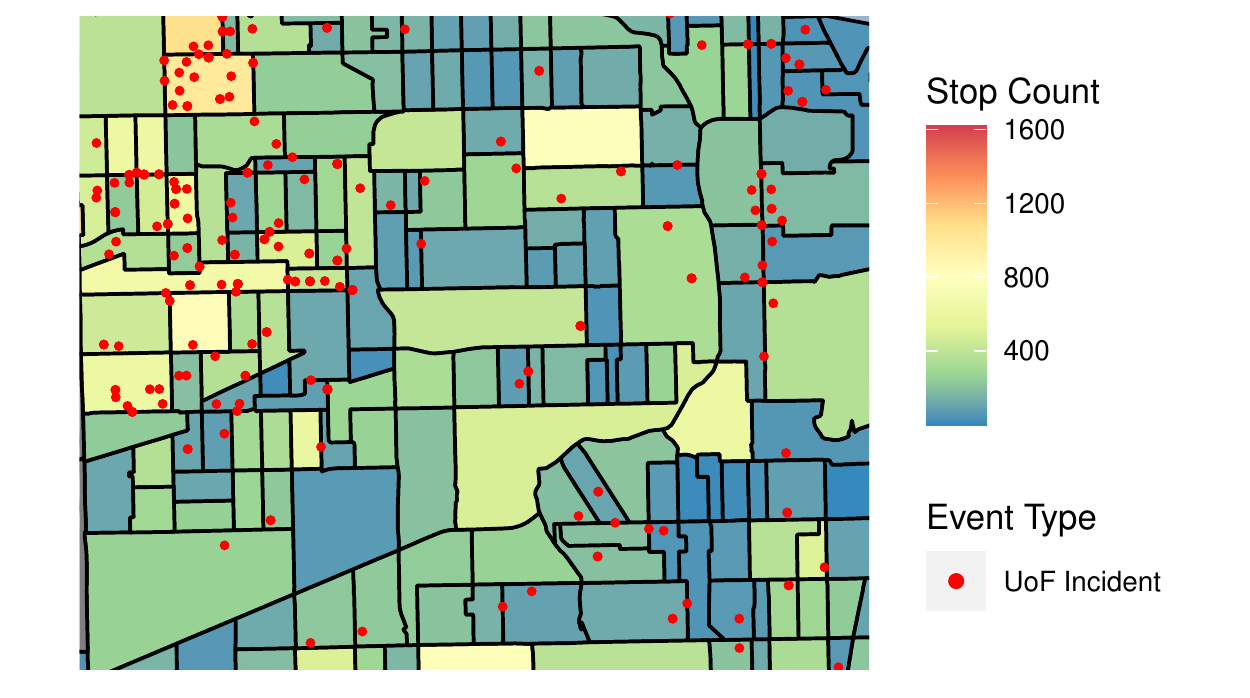}
    \caption{Use of Force Incidents with Counts of Police Stops by Census Tract, Chicago}
    \label{fig:stop_count_areal}
\end{figure}

When studying the relationship between violent crime and use of force, the spatial level of analysis for violent crime often varies between studies. Many studies use neighborhood-level measures of violent crime, such as the number of violent crimes or homicides per police district, command area, or city per a certain number of residents \citep{reisig2003neighborhood, lawton2007levels, lee2010examination, terrill2003neighborhood}. This method is depicted for police stops in Chicago in Figure \ref{fig:stop_count_areal}, where each use of force incident is affiliated with the count of police stops per census tract. No information about the police stop is preserved, other than the census tract where it is located.

\begin{figure}[ht]
    \centering
    \includegraphics[width = 0.75\textwidth]{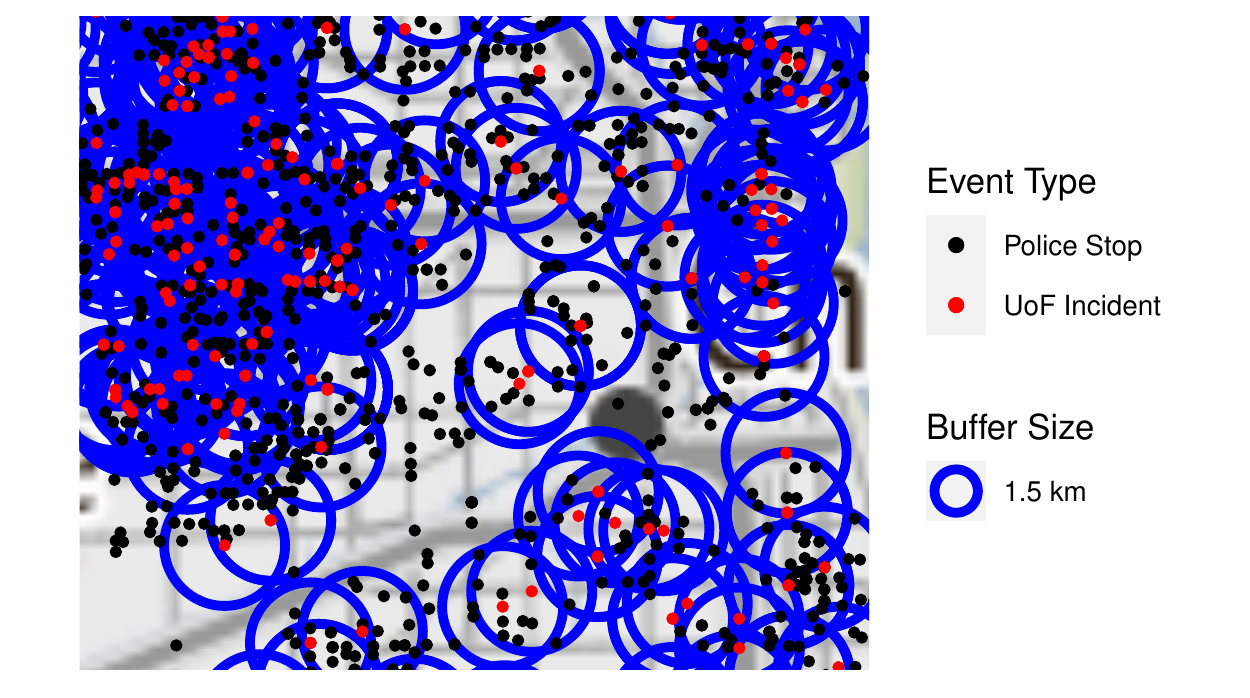}
    \caption{Illustration of Radial Buffers for Chicago}
    \label{fig:buff_zone}
\end{figure}

Neighborhood-level violent crime rates have limitations, as neighborhoods are not spatially homogeneous, which suggests the importance of analysis at lower levels of aggregation. %
To address these limitations, \cite{lee2014impact} uses radial buffers to count the violent crimes within a certain radius of each use of force incident. The radial buffers range from 500 to 3,000 feet. These counts of violent crimes within a certain radius are then used as a covariate in the model, instead of a count per neighborhood. We illustrate this general method in Figure \ref{fig:buff_zone} where we show a subset of use of force incidents in a part of Chicago in red. The blue radial buffers are used to count the number of police stops (black) that occur within 1.5 km, in this case, from each use of force incident. 

\cite{lee2014impact} runs four separate multinomial logistic regression models at the four levels of radial buffers and finds that the micro-level use of place (using the radial buffers) is important when modeling the relationship between violent crime and police use of force. The results indicate that smaller radial buffers lead to a larger positive relationship between violent crime and police use of force and that larger radial buffers lead to a possible `blur' of the genuine effect of violent crime on police use of force. The findings from this work show that the relationship may be lost almost altogether when looking at the neighborhood level. In total, this suggests that using the smallest level of aggregation possible decreases the potential of losing information.

Radial buffers preserve some of the granularity of the violent crime/police stop locations when determining their relationship with police use of force. We can create buffers that are smaller than census tracts and the sizes of all buffers could be uniform, as in \cite{lee2014impact}, or they could vary based on variables such as population density.  However, this method is still a form of aggregation, and therefore we lose many of the precise details about the violent crime incidents or police stops. For example, we have to fix the radial buffer, so after the size is set we cannot know how many crimes occur within both 10 feet and 500 feet of the given use of force incident. We also cannot preserve precise point or event-level information about the violent crime incidents or police stops. These could be variables such as the race and gender of the officer and/or civilian, if a gang was involved, if a weapon was involved, and if the incident was a hate crime. When we aggregate to radial buffers, we no longer have the ability to use event-level information, such as these, for the violent crime incidents or police stops in the point process model.

\section{Methods}
\label{section_methods}

We develop a spatial point process approach as follows. Let the $n$ random locations of police use of force incidents $x_1,\dots, x_n \in D \subset {\mathbb R^2}$ be a point process with intensity function $\lambda_1(s), s\in D$. $D$ is the study region, for instance the city of Chicago. The $m$ random locations $y_1,\dots, y_m\in D$ of the second point process, police stops in our application, can be related to the first point process using multiple methods. %
In case-control models this point process has baseline intensity function $\lambda_0(s)$ %
 while in our shared component model, it has a second intensity function $\lambda_2(s)$. In our shared component framework, we relate both point processes to spatial covariates, $\bm{z(s)}$, through regression coefficients $\bm{\beta}$. In case-control models, only the case process,  use of force incidents in our data, are related to spatial covariates.

Our aim is to relate two or more point processes while preserving the point-level information of all datasets. In what follows, we describe relevant models in the literature, mostly focusing on marked point process models and summary or test statistics. Next, we present two classes of models that we explore in detail, the first being the well-established case-control model and the second being a novel use of a shared component model for point processes. We describe a Bayesian inferential approach for relating two point processes for both classes of models.

\subsection{Existing Methods to Relate Multiple or Multitype Point Processes}

Multitype point processes, where the marks are categorical labels, offer one approach to relating two or more point processes. 
 In such point processes, the mark determines the type of point at a given location. The relationship between types of points, determined by the categorical mark, is often modeled through a cross-covariance function in a bivariate point process or tests of similarities of first-order characteristics. For example, \cite{fuentes2017nonparametric} proposes a nonparametric comparison of multitype point processes based on first-order properties, which describe the spatial distribution in the area of interest. The methods are used to analyze different types of wildfires in Spain based on two characteristics: size (small, regular, large) and cause (arson, natural, negligence, reproductions, and unknown).  \cite{berman1986testing} tests the relationship between a spatial point process and other spatial stochastic processes. Innovative spatial data types are used with the goal of analyzing the relationship between copper deposits (a point process) and linear features observed in the region, called `lineaments'; these are often roads or perceived breaks in the earth's crust, as observed by a satellite. \cite{berman1986testing} develops a test statistic for testing the relationship between the two processes based on the distances from the points of the point process (copper deposits) to the nearest point in the stochastic process (lineament). %
The tests proposed by \cite{fuentes2017nonparametric} and \cite{berman1986testing} are similar to other recently developed multitype test statistics \citep[cf.][]{illian2008statistical, moller2003statistical}. These tests are useful exploratory tools and can motivate further exploration through parametric models of the intensity of point processes.

\cite{mohler2014marked} uses marked point process models to create hotspot maps. The categorical mark for this point process framework specifies different crime types, including homicide and crime types that may precede homicide. \cite{mohler2014marked} determines the probability that an event $i$ triggered a given homicide event, where event $i$ could be another homicide or a different crime type, through a self-exciting point process framework. All of the crimes that are not homicides involve a handgun; these data are used to analyze the relationship between gun crimes and the prediction of future homicide. This innovative space-time approach incorporates two or more marks, which is useful when considering many marks. The approach relies on short and long-term kernel density estimates whereas we pursue a parametric approach to evaluate the relationship between two point processes. A triggering function may be a useful tool in future analysis to incorporate the temporal dimension of both point processes.

\cite{liang2008analysis} develops a bivariate mark point process model that allows for dependence across levels of a mark within the point process model. The model includes spatial variables and their regression coefficients, $\bm{z(s)}$ and $\bm{\beta}$, nonspatial variables and their regression coefficients, $\bm{\nu}$ and $\bm{\alpha}$, and a Gaussian process $\omega(s)$.  The log spatial intensity function for mark $k$ of the point process is defined as $\log(\lambda_{\bm{\theta}_k}(s, \nu)) = \bm{z(s)}'\bm{\beta_k} +\bm{\nu}'\bm{\alpha_k} + \omega_k(s)$. The mark considered by \cite{liang2008analysis} is cancer type, where two cancer types are considered. This model allows flexible spatial dependence across two mark levels, $k=1,2$, through a cross-covariance function for the Gaussian processes, $\omega_k(s)$. \cite{kelling2021two} find that interpretation is difficult due to the use of nonspatial variables in the spatial intensity function. Furthermore, analysis of the dependence between the two mark levels is limited to the $\rho$ parameter of the dependent Gaussian processes, with cross-covariance function $\Lambda$, defined below. The parameter $\rho$ gives information on the strength of the dependence in the spatial residual pattern, but we aim to analyze the relationship between two point processes in more detail. %

$$
\Lambda = 
\begin{bmatrix}
\sigma_{1}^2 & \rho\sigma_1\sigma_2\\
\rho\sigma_1\sigma_2 & \sigma_{2}^2.\\
\end{bmatrix}
\quad
$$

\subsection{Case-Control for Point Processes}

\label{sec:case_cont_desc}

Case-control methods represent a common class of models used to compare two point processes. \cite{guan2008second} use case-control point process methods to relate a 1990 survey of birds (the baseline/control process) to a more recent survey in 2004 (the main/case process). The results show that the spatial distribution of golden plovers have changed over time in relation to spatial covariates, namely slope and cotton grass coverage. \cite{diggle2007second} use case-control methods to study the relationship between juvenile and adult trees in a tropical rain forest. Juvenile trees are treated as the controls, which represent underlying environmental variation, and the adults are the cases whose spatial distribution is impacted by spatial conditions affecting survival, such as elevation. \cite{guan2010nonparametric} also study the relationship between trees in tropical rain forests, where dead juvenile trees are treated as cases and new trees are treated as controls in order to study the mortality of juvenile trees. Spatial covariates used in this analysis include altitude, slope, and information on the soil content. The second example studied by \cite{guan2010nonparametric} includes the golden plover data from \cite{guan2008second} with spatial variables altitude, slope, and percent cover of heather and cotton grass. \cite{diggle2000point} and \cite{chetwynd2001investigation} extend case-control methods to matched case-control data, where a set of matched controls are assigned to each case based on potentially relevant confounders.

To introduce the notation for case-control methods, the intensity function for the main point process of interest at location $s$, called the case process, is denoted $\lambda_1(s)$. The intensity for the baseline process at location $s$, or the control, is denoted $\lambda_0(s)$. Spatial variables are denoted $\bm{z(s)}$ with corresponding regression coefficients $\bm{\beta}$ and the intercept term is denoted $\alpha$. The full intensity function for location $s$ used in case-control point processes is as follows \citep[cf.][]{diggle2007second}: 

\begin{equation}
    \lambda_1(s) = \lambda_0(s)\exp\left(\alpha + \bm{z(s)}'\bm{\beta}\right)
    \label{case_cont_eq}
\end{equation}

A slightly different parameterization is often used where the intercept term scales the overall intensity function as follows: $\lambda_1(s) = \alpha\lambda_0(s)\exp\left(\bm{z(s)}'\bm{\beta}\right) $ \citep[cf.][]{diggle1990point, diggle1994conditional, guan2008second}. In this formulation, $\lambda_0(s)$ is the intensity of the population at risk and $\alpha$ is a scaling parameter which represents the prevalence of the cases relative to controls
and is often not considered to be of primary scientific interest \citep{diggle1994conditional}. The remaining portion of $\lambda_1(s)$, $\exp\left(\bm{z(s)}'\bm{\beta}\right)$, determines the elevated risk of the cases as a function of spatial variables. In these case-control models, no parametric form is assumed for the baseline intensity $\lambda_0(s).$

In many cases direct parametric estimation of the full spatial intensity function of the case point process $\lambda_1(s)$ is avoided because of complications introduced through estimation of the baseline intensity $\lambda_0(s)$.
\cite{guan2008second} estimates the pair correlation function for the main/case process in order to characterize the second-order structure, for example clustering, of the main process after accounting for the baseline process. \cite{guan2010nonparametric} develop both nonparametric and parametric ways to study the second-order structure of the main point process. Methods are proposed to estimate regression coefficients for the main process without estimation of the control intensity $\lambda_0(s)$. The simulation study fixes regression coefficients, $\bm{\beta}$, but does not evaluate the ability of the estimation procedure in recovering the regression parameters. Rather, the simulation study is focused on the bias and standard deviation of the estimators for the pair correlation function in order to evaluate clustering. We also note that independence is assumed between the case and the control process. \cite{diggle1994conditional} also avoids estimation of $\lambda_0(s)$ through the use a non-linear binary regression model to estimate $\bm{\beta}$.

Some case-control methods pursue nonparametric estimation of the baseline intensity function $\lambda_0(s)$ through kernel density estimates which are then used to estimate the intensity of the case process, $\lambda_1(s)$ \citep{diggle2000point, diggle2007second}. This is useful when we would like to visualize and interpret the full spatial intensity function, $\lambda_1(s)$. \cite{diggle2007second} proposes a simple estimation procedure for the regression coefficients in the case-control model without estimation of the control intensity, $\lambda_0(s)$. Coefficients $\alpha$ and $\bm{\beta}$ are estimated through logistic regression where $Y(s)$ is binary and indicates whether a given location $s$ is a case (1) or control (0). The logistic regression takes the following form: $ P(Y(s) =1 |\bm{z(s)}) = \frac{\exp(\alpha + \bm{z(s)'\beta})}{1+\exp(\alpha+\bm{z(s)'\beta})} $. If it is desired to estimate the case intensity, $\lambda_1(s)$, the control intensity $\lambda_0$ is estimated through a kernel density estimate. The results from logistic regression and the kernel density estimate combined create the estimate of the intensity function $\lambda_1(s)$. \cite{diggle2007second} interprets regression coefficients $\bm{\beta}$ as the effects of $\bm{z(s)}$ on the relative intensity of cases to controls at location $s$. The intercept $\alpha$ is interpreted as the ``chosen ratio'' between cases and controls. %

\cite{hessellund2021semiparametric} considers a multivariate extension of the case-control logistic regression model illustrated in \cite{diggle2007second}. Importantly, this multivariate case-control model does not assume independence between the various point processes. \cite{hessellund2021semiparametric} analyzes six types of street crimes in Washington DC: Robbery, Auto Theft, Vehicle Theft, Assault, and Burglary, and Other Theft. The intensity for each point pattern $X_i$ is defined as $\lambda_i({\bm u}; {\bm\gamma_i}) = \lambda_0({\bm u})\exp\left({\bm\gamma_i}'{\bm z(u)}\right), i = 1,...,p$. Spatial covariates ${\bm z(u)}$ are defined at location ${\bm u}$ with regression coefficients ${\bm\gamma_i}$. The background intensity is defined as $\lambda_0({\bm u})$ and is interpreted as the ``spatial effects of latent factors such as the urban structure and population density and is assumed to be common for all point types'' \citep{hessellund2021semiparametric}. The parameters $\bf{\gamma_i}$ are shown to be not identifiable so regression coefficients ${\bm\beta_i}$ are defined so that ${\bm\beta_i} = {\bm\gamma_i} - {\bm\gamma_p}, i = 1,...p-1$ where $X_p$ is the baseline point process. The regression coefficients for the baseline point process, ${\bm\gamma_p}$ are set to 0 so that $\lambda_0({\bm u})$ is the intensity of the baseline process. In the application to the Washington DC street crimes, the `Other Theft' category of crimes is set to be the baseline category and all other categories are compared to this point process.

\cite{hessellund2021semiparametric} once again avoids estimation of the baseline intensity $\lambda_0(s)$ when estimating regression coefficients ${\bm\beta}$ through the use of multinomial logistic regression, similar to that of \cite{diggle2007second}, which does not depend on $\lambda_0(s)$. For interpretation of the Washington DC street crimes, \cite{hessellund2021semiparametric} plots conditional probability maps based on this multinomial logistic regression. In order to plot $\lambda_i(s)$, the authors then estimate $\lambda_0(s)$ through a kernel estimate. %

\cite{xu2019stochastic} introduce a modified version of the case-control model where a spatially varying control process is scaled by a sampling scheme, $\alpha(s)$, which is often assumed to be known. For the analysis of restaurants in Beijing in \cite{xu2019stochastic}, a uniform 6\% sampling rate is assumed for the controls. The intensity for controls, $\lambda_M(s) = \alpha(s)\lambda_0(s)$, and the intensity for cases, $\lambda_N(s) = \lambda_0(s)\exp\left(\bm{z(s)'\beta}\right)$, are both dependent on the process $\lambda_0(s)$. Once again, the estimation of $\lambda_0(s)$ is avoided due to the strategic use of the proportional intensity functions to estimate regression coefficients. The term $\lambda_0(s)$ is denoted an ``infinite-dimensional nuisance parameter" due to concerns over inconsistent estimation of $\lambda_0(s)$ and the effect on inference for regression coefficients. Independence is assumed between the case and control processes.%

In our analysis of simulated examples and policing data from Chicago, we consider two methods of estimation for the case-control model: logistic regression and as a spatial intensity function for a nonhomogeneous Poisson process. All models in our paper are implemented through a Bayesian hierarchical framework, except the case of logistic regression, which is estimated through maximum likelihood and the \textit{glm} function in R.

\subsection{Shared Component Model for Point Processes}
%

%

We develop a shared component model for point processes, building upon the areal data framework in \cite{knorr2001shared}. Similar to case-control models, we scale the intensity function by a term, $\lambda(s)$. However, in our case, this shared component is weighted by the parameter $\delta$ and contributes to both point processes, rather than representing the intensity of one point process. This shared component is inferred from the data, rather than calculated from nonparametric methods or ignored, as is often the case with the control process in case-control methods. \cite{knorr2001shared} study the spatial distribution of two different types of cancer, oral cavity and oesophagus, which have been shown to have common and unique risk factors. Similarly, we would like to infer the shared spatial pattern between multiple types of policing data, including police use of force, police stops, and violent crime incidents.

Our shared component model is shown in Equation \ref{shared_comp_eq}, where $\lambda_1(s)$ denotes the spatial intensity of one point process, and $\lambda_2(s)$ denotes the spatial intensity of the second point process. The spatial variables for each point process, $\bm{z(s)}$, and their corresponding regression coefficients, $\bm{\beta_1}$ and $\bm{\beta_2}$ allow us to gain knowledge about the spatial pattern that is unique to each individual spatial pattern, rather than shared between them. For example, we may assume that population effects may largely be captured by the shared component, $\lambda(s)$, while other neighborhood effects may be unique to each point process. From our simulation studies, we have found that we can include the same spatial variables in both intensity functions and recover the corresponding regression coefficients accurately, although this should be tested in more detail for each application and context.

\begin{equation}
    \begin{array}{l}
    \lambda_1(s) = \lambda(s)^\delta \exp\left(\bm{z(s)}'\bm{\beta_1} \right) \\ \newline
     \lambda_2(s) = \lambda(s)^{1-\delta} \exp\left(\bm{z(s)}'\bm{\beta_2}\right)  
    \end{array}
    \label{shared_comp_eq}
\end{equation}

There are many variations that could be considered of the shared component model presented in Equation \ref{shared_comp_eq}. The simplest version of the model is to consider a shared component that does not vary over space, or $\lambda(s) = \lambda$ for all $s$. Through simulation studies, we have found that in this case we cannot include intercept terms in the intensity for either point process, as they are confounded with the shared component which is essentially a shared intercept term between the two point processes. For our analysis, we focus on a spatially varying shared component, where $\log(\lambda(s))$ is a Gaussian process. We estimate the Gaussian process using predictive process transformations due to the computational burden of estimating an $n\times n$ covariance matrix \citep[cf.][]{banerjee2008gaussian}. We estimate the Gaussian process over a small number of knots and then transform this Gaussian process to the data points using the covariance function between the knots and the data points. For our study, we use 82 knots evenly distributed over the region, as shown in Figure \ref{fig:integ_predproc_points}. The Gaussian process over the knots ($\lambda^*(s)$) is transformed to the estimate of the Gaussian process over the data points ($\Tilde{\lambda}(s)$) using the covariance function between the knots and between the points and the knots, as shown below. %

\vspace{-12pt}

     $$\Tilde{\lambda}(s) =  \text{cov}(\lambda(s), {\lambda^*(s)})\text{var}^{-1}\left({\lambda^*(s)}\right){\lambda^*(s)}$$

\vspace{-2pt}

There are also multiple choices for the parameterization of the weighting parameter, $\delta$. We focus our studies on two parameterizations: weights of $\delta$ and $1-\delta$ for each point process, as shown in Equation \ref{shared_comp_eq}, and weights of $\delta$ and $1/\delta$ for the two point processes, respectively. In the first case, $\delta$ is bounded between 0 and 1 and we use a Uniform(0,1) prior for $\delta$. Other choices of bounded distributions, such as a Beta distribution, could also be used. In the second case, $\delta$ is not bounded between 0 and 1. The log of $\delta$ has a Normal prior distribution, as suggested by \cite{knorr2001shared}. Overall, we find that the parameterization with the shared component weights being $\delta$ and $1-\delta$ produce reliable parameter estimates and easier interpretation than the second point process being weighted as $1/\delta$. We note that we have constrained $\delta$ in both examples so that the shared component has a positive contribution to both point processes ($\geq 0$). Although the shared component $\lambda(s)$ must be positive, other parameterizations of $\delta$ could include a negative contribution of the shared component to the point process intensity function. %

The likelihood for the shared component model is included below. To estimate this likelihood, we must estimate the integral of the individual point process intensity functions, $\lambda_1(s)$ and $\lambda_2(s)$, over the region of interest, $W$. We estimate this integral using Monte Carlo averages of the values of each intensity function over integration points, described in Section \ref{sec:integ_methods_app}. The intensity functions for both point processes are then multiplied over all of the data points for each point process ($n$ points of the first point process, $m$ points of the second point process). %

\vspace{-12pt}

$$\mathcal{L}(\bm{\theta}; \{s_{1:n}\}, \{s_{1:m}\}, s_i\in W) \propto  \exp\left(-{\int_{W} \lambda_{1}(s)ds - \int_{W} \lambda_{2}(s)ds}\right) \times \prod_{i=1}^{n} \lambda_{1}(s_{i}) \times \prod_{j=1}^{m} \lambda_{2}(s_{j})  $$

\vspace{-2pt}

Finally, we could consider different parameterizations of the part of the intensity function that is specific to each point process. For example, we could consider different combinations of spatial variables across both intensity functions, including different or identical variables between point processes. This is an advantage of our shared component method; case-control methods do not incorporate parametric estimation of the intensity of the control process. In our parameterization shown in Equation \ref{shared_comp_eq}, we also show a nonhomogeneous Poisson process (NHPP) for both point processes for the part of the intensity function that is not shared between both point processes. Other forms of this intensity function could be explored in the future, such as forms of a log Gaussian Cox process. In this case, care must be taken to avoid confounding between the shared component and the Gaussian process specific to the point processes. We note that in our framework of NHPP's for both frameworks, we find the most reliable estimation when an intercept term, $\beta_0$, is only included for one of the point processes, not both.

Our shared component model does not assume one process is a ``baseline process" whereas the other process is the ``case" or ``main" process. This has many advantages, especially when the direction of causality is not known. In regards to policing, for example, many studies shown above have said that violent crime provides context of neighborhoods, which may impact policing behavior. However, there are also policing studies, such as the Broken Windows theory, that posit policing behavior can also impact the prevalence of violent crime. Specifically, the Broken Window theory states that if ``soft'' crimes are tolerated by police, more criminals may commit crimes in that given area \citep{wilson1982broken}. Increased police presence in one neighborhood has also been associated with the spatial displacement of crimes to other areas \citep[cf.][]{ratcliffe2002burglary}. Given that policing behavior may affect crime and crime may affect policing behavior, our shared component model is advantageous in this analysis.
We note that the shared component model can be written as a version of the case-control model where the scaling process now depends on the intensity function of the second process, the parameter $\delta$ defined above, as well as the spatial covariates and coefficients for the second point process. This is discussed in more detail in Section \ref{sec:app_case_cont_shared}.

We describe two interesting case-control based approaches and compare their work to the shared component model that we develop. In the example of tropical rain forest data provided by \cite{xu2019stochastic}, there is not an apparent control process available to study the three tree species. Therefore, a homogeneous Poisson process is used as the control process and all three tree species are treated as cases compared to the uniform control. \cite{xu2019stochastic} note that the size of the homogeneous Poisson process, determined by a varying $\alpha$ where $\lambda_0(s) =1$, actually has an effect on coefficient estimation for the case processes. The shared component model that we develop here avoids specifying one process as the control process. In the simulated case-control example presented by \cite{guan2010nonparametric}, the authors assume the control intensity takes the following form: $\log \lambda_0(s) = \log \eta + X^*(s)$. The case process intensity is assumed to take the following form: $\log \lambda_1(s) = \log \gamma + \bm{\beta} X(s) + X^*(s).$ The process $X(s)$ is assumed to be known, while $X^*(s)$ is assumed to be unknown. The term $X^*(s)$ is similar to an unweighted shared component between the two point processes from our shared component framework.

\section{Data}
\label{section_data}

We evaluate the case-control and shared component methods through simulated data as well as policing datasets from Chicago. In this section, we discuss the generation of our simulated datasets and the policing datasets from Chicago.

\subsection{Simulated Data}
\label{sec:sim_data}

\begin{figure}[H]
    \centering
    \includegraphics[width=0.97\textwidth]{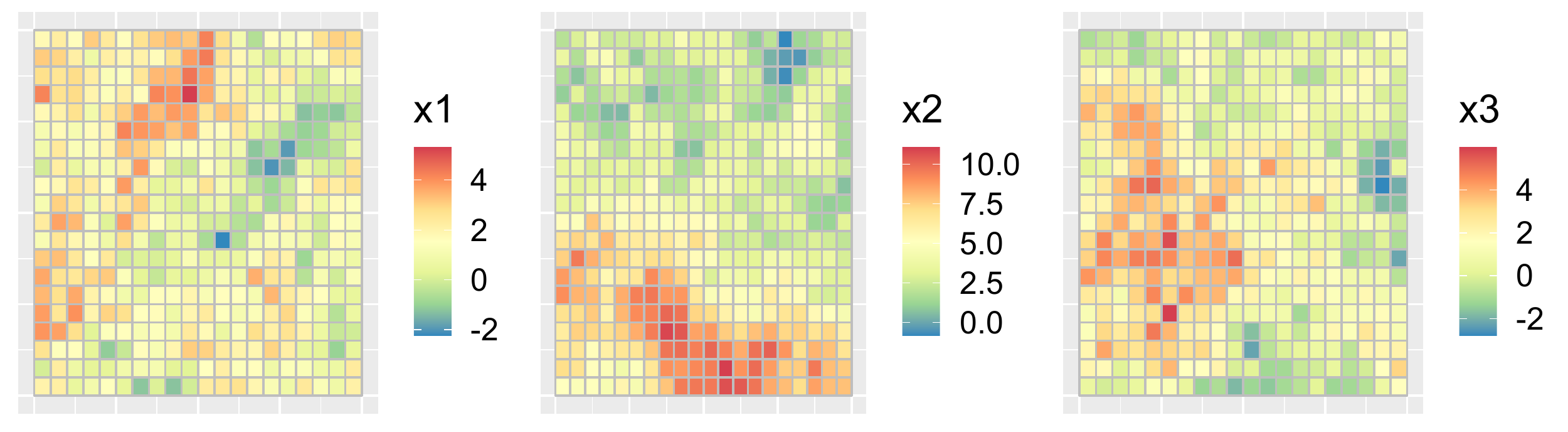}
    \caption{We simulate three spatial variables on a lattice over the unit square that will then be used to simulate point processes under the case-control model and the shared component model. The lattice is a simple case of areal data, which will be extended to census tracts for the case of Chicago. Different combinations of these spatial variables are used to determine the spatial distribution of one or both point processes in the simulated examples.}
    \label{fig:sim_spat_var}
\end{figure}

We use the process of spatial thinning to generate all of the simulated point processes in this paper. First, we simulate a homogeneous Poisson process over the spatial window corresponding to the area of the window and the maximum possible intensity over the spatial window. The point process is then thinned and the probability of keeping a point is equal to the intensity at that point divided by the maximum intensity in the region. The spatial intensity functions are dependent on one or more spatial variables, which are depicted in Figure \ref{fig:sim_spat_var}. All spatial variables and point processes are simulated over the unit square.

For case-control models, we simulate a baseline/control process using a parametric intensity function, in order to give the point process a nonhomogeneous distribution over space. We denote this parametric intensity function $\lambda_0(s)$ in Equation \ref{case_cont_eq}. The intensity of the first point process, $\lambda_0(s)$, and spatial variables are used to simulate a second point process based on its full spatial intensity $\lambda_1(s)$, as shown in Equation \ref{case_cont_eq}. Importantly, we simulate the case-control data using a spatial intensity model specification, not using logistic regression, so we expect the spatial intensity function to perform well when recovering parameters. %

For the shared component models, the two point processes are simulated simultaneously instead of sequentially. First, we simulate a Gaussian process to serve as the log of the shared component, so that the shared component is always greater than zero. We denote $\log(\lambda(s)) \sim GP(0, \Sigma)$ where $\Sigma$ is a univariate exponential covariance function such that the covariance for two locations $s_i, s_j$ is defined as $\Sigma = \sigma^2 \exp\left(\frac{-|s_i-s_j|}{\phi}\right) $, with $\sigma^2, \phi > 0$. The $\phi$ parameter is fixed in our simulation studies and real data application and is not estimated from the data, as is common in other studies due to identifiability issues \citep[cf.][]{liang2008analysis}. We calculate the value of $\phi$ so that the 95$^{th}$ percentile of distances between all points would have a a correlation of 0.05, and the value of $\phi$ so that the 5$^{th}$ percentile of distances would have a a correlation of 0.95. We fix $\phi$ at the average of these two values. %
The parameter $\sigma$ is also fixed in simulation studies but then is treated as unknown when estimating model parameters. After simulating the Gaussian process, we fix the weight parameter $\delta$, which decides the contribution of the shared component to each of the two point processes, as well as the regression coefficients for each point process. Finally, we use spatial thinning to generate the two point processes using their corresponding spatial intensity functions, as shown in Equation \ref{shared_comp_eq}.

\subsection{Chicago Policing Data}

Next, we describe the police use of force and police stop datasets used in this study. \cite{ba2021role} acquired detailed data from the Chicago Police Department through the use of open-records requests and appeals. We utilize the replication data posted through the Code Ocean capsule \citep{ba2021role_code}. Both the use of force data and the police stop data cover 2012-2015. We note that there is a large number of events that do not have coordinates (latitude/longitude) given for the events. Specifically, 18.8\% of use of force events and 14.6\% of police stops do not have coordinates available. In this analysis, we remove these points, but future work may require additional investigation into the missingness of these points.

\begin{figure}
    \centering
    \includegraphics[width=0.97\textwidth]{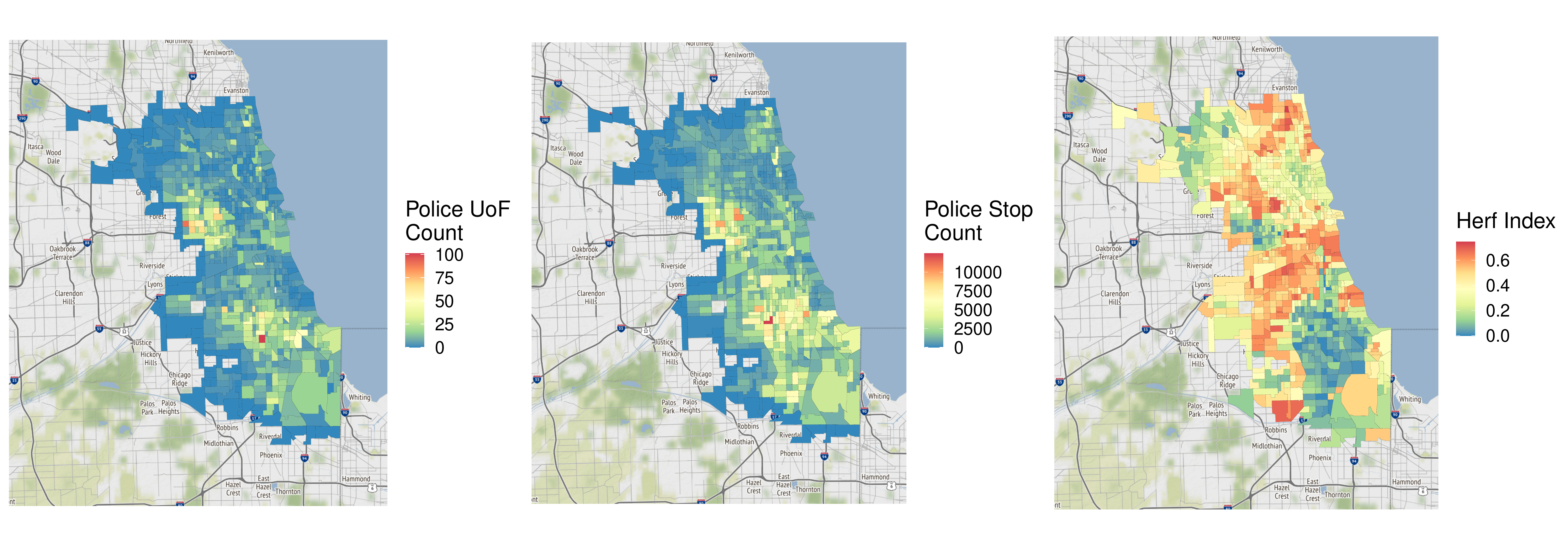}
    \caption{Police Use of Force Counts, Stop Counts, and Herfindahl Index by Census Tracts}
    \label{fig:chic_spat_var}
\end{figure}

The use of force dataset from the Chicago Police Department includes 9,293 incidents from January 2012 through December 2015, 7,539 of which have complete location information. The data also includes information about individuals involved in the event such as civilian race, gender, age, and injury status as well as the officer ID. The dataset of police stops includes 1,703,158 incidents from January 2012 through December 2015, 1,453,832 of which have complete spatial information. The dataset also includes information such as the type of stop, the civilian race, gender, and age, and the officer ID. We plot both datasets in Figure \ref{fig:chic_spat_var} on the areal level, for ease of visualization. We notice that the two outcomes share a spatial pattern, where there are smaller counts on the border of the city and larger counts in the southern center of the city. Note that we have only plotted this data on the areal level for ease of visualization for hundreds of thousands of points- our analysis is on the point-level in continuous space.

In addition to the two policing datasets analyzed for Chicago, we also collect socioeconomic information from the US Census and the American Communities Survey. We use the census tracts that are completely contained within the police beats, the latter of which were downloaded from Chicago's Open Data Portal. We focus our attention on three socioeconomic variables gathered or calculated from the census data: median age, unemployment rate, and the Herfindahl Index for neighborhood diversity. We plot the Herfindahl Index for Chicago census tracts in Figure \ref{fig:chic_spat_var}. We note that the models we introduce here are flexible to difference choices of covariates, depending on the application and research questions.

\section{Results}
\label{section_results}

We compare the case-control and shared component models using both simulated datasets and real policing data from Chicago. We test many different parameter settings for the case-control model to illustrate that commonly used estimation methods should be interpreted with caution. We also apply the shared component method to simulated data and policing data from Chicago and find that it provides a flexible framework to analyzing the relationship between two point processes. 

For estimation of all parameters, we use Markov Chain Monte Carlo (MCMC) implemented through the NIMBLE package in R \citep{nimble}. Our approach relies on Metropolis-Hastings adaptive random-walk samplers with univariate normal proposal distributions for the regression coefficients (Normal(0,100)) \citep{nimble}.  We use an Inverse-Gamma($\alpha = 2$, $\beta = 0.5$) prior for $\sigma$, the parameter associated with the covariance function of the shared component, as in \cite{liang2008analysis}. We assess convergence through trace plots, effective sample size \citep{gong2016practical}, and Monte Carlo standard error \citep{batchmeans_package, mcmcse_package}.

\subsection{Results for Simulated Data}

Through simulation studies, we have found that the case-control model presents some challenges with interpretation depending on model assumptions. We simulate from the case-control model as a spatial intensity function with many different simulated parameter settings for the regression coefficients. To do this, first we simulate data from one point process. We proceed to simulate a second nonhomogeneous Poisson process (NHPP) based on the spatial intensity function described in Equation \ref{case_cont_eq}, using the true intensity function for the baseline point process, $\lambda_0(s)$. In practice, we often do not know the functional form for $\lambda_0(s)$ so we must estimate it using a kernel density estimate (KDE) or avoid its estimation altogether. We discuss two possible estimation procedures below.

After simulating the case and control NHPPs from the spatial intensity functions corresponding to the case-control model, we use two estimation methods, the spatial NHPP and logistic regression, to test parameter recovery for the case intensity function. In the first case, we estimate the spatial NHPP parameters through a Bayesian approach with the following likelihood function, where $\lambda_1(s)$ is the case intensity function, which includes regression coefficients $\alpha$ and $\bm{\beta}$ and the control intensity, $\lambda_0(s)$. We use independent mean 0 normal priors for the regression coefficients. We do not assume any knowledge of the baseline intensity function, and therefore use a KDE estimate of the control intensity function, $\hat{\lambda_0(s)}$, when estimating the case intensity function parameters.

$$\mathcal{L}(\alpha, \bm{\beta}; s_1,...,s_n, s\in W) \propto  \exp\left(-\int_{W} \lambda_1(s)ds\right) \times \prod_{i=1}^n \lambda_1(s_{i}) $$

For the second approach, we use logistic regression as described in Section \ref{sec:case_cont_desc}. This approach avoids the use of the baseline intensity function, $\lambda_0(s)$, when estimating regression coefficients. If these two methods of estimating parameters from the same data result in similar parameter estimates, then we have flexibility between the estimation methods and corresponding interpretation. 

As shown in Table \ref{tab:case_cont_sim_unit}, we find that when all parameters are positive, the logistic regression approach is able to recover regression parameters well. There are challenges when using the NHPP approach to estimate regression parameters, due to the use of a KDE estimate of the baseline intensity function. When we include negative parameters, as shown in the second simulated example, we are not able to recover the intercept parameter well with logistic regression and some other parameter estimates are also impacted. We test the sensitivity of these results to scaling (for example, using an intercept parameter of -1 instead of -10) and still find challenges in estimating $\beta_0$ using logistic regression. We include the credible intervals for the NHPP estimated using MCMC and confidence intervals for the logistic regression, estimated through the \textit{glm} function in R. From these results based on simulated data, logistic regression presents a promising alternative to estimation with a NHPP but can still present challenges in some settings. %

\begin{table}
\caption{\label{tab:case_cont_sim_unit} Estimated regression coefficients with confidence/credible intervals (logistic regression/NHPP, respectively) for simulated datasets on the unit square.} \\
\centering
\adjustbox{max width=1\columnwidth}{
\begin{tabular}{|c|ccc|}
\hline
         Model & $\beta_0$ & $\beta_1$ ($x_1$) & $\beta_2$ ($x_2$) \\
         \hline
         \hline
         Truth & -10.0  &  1.7 &  0.8\\
         NHPP est. (int.) &  -12.93 (-13.17, -12.70) &  1.75 (1.71, 1.79)  &  0.84 (0.82, 0.86)  \\
         Log Reg est. (int.) & -9.84 (-10.26, -9.42) &   1.70 (1.63, 1.79)  &   0.78 (0.74, 0.82) \\
         \hline
         Truth & -10.0 &  2.1 & -0.1 \\
         NHPP est. (int.) & -12.40 (-12.59, -12.22) & 2.08 (2.05, 2.12)  & -0.10 (-0.12, -0.09)  \\
         Log Reg est. (int.) & -5.29 (-5.58, -5.01) &  2.14 (2.06, 2.23)  &  -0.14 (-0.16, -0.11) \\
         \hline
\end{tabular}
}
\end{table}

In a second set of simulated examples, we assume we have some knowledge of the baseline intensity function and test if the NHPP method can accurately estimate parameters with these assumptions. We simulated data from the control process with intensity function $\lambda_0 = \exp(\beta_{0,\text{control}} + \beta_{1,\text{control}} x_3)$ and the case process with intensity function $ \lambda_1(s) = {\lambda}_0(s) \exp(\beta_0 + \beta_1x_1 + \beta_2x_2)$. We assume that we know the form of the baseline intensity function and estimate the regression coefficients corresponding to both the case and the control intensities. We note that the intercepts are not identifiable and we estimate the sum of the intercept parameters for both point processes. In Table \ref{tab:case_cont_sim_unit_intknown}, we show that when we assume we know some information about the structure of the baseline intensity, we are able to estimate the coefficients accurately using the NHPP approach. In practice, specifying a structure for the baseline intensity may not be possible.

\begin{table}
\caption{\label{tab:case_cont_sim_unit_intknown} Estimated regression coefficients with credible intervals for simulated datasets on the unit square when $\lambda_0(s)$ parameters are estimated.} 
\centering
\adjustbox{max width=1\columnwidth}{
\begin{tabular}{|c|cccc|}
\hline
         Model & $\beta_0 + \beta_{0,\text{control}}$ & $\beta_1$ ($x_1$) & $\beta_2$ ($x_2$) & $\beta_{1,\text{control}}$ \\
         \hline
         \hline
         Truth & -5.8  &  2.1 &  -0.1 & 0.7 \\
         NHPP est. (int.) & -5.75 (-5.96, -5.54) &  2.09 (2.04, 2.13)  &  -0.10 (-0.11, -0.08) & 0.69 (0.66,0.71) \\
         \hline
\end{tabular}
}
\end{table}

From these findings, we suggest caution when interpreting results from both the NHPP and logistic regression estimation methods \cite{diggle2007second} if the assumption is that these results should be interpreted as a spatial intensity function. We find that these two estimation procedures can produce different results, so it is important to consider the desired interpretation of these parameters. Logistic regression has many advantages, including avoidance of estimating the baseline intensity function $\lambda_0(s)$ when estimating regression coefficients and computationally efficient estimation. We have shown some advantages of this approach when compared to the full Bayesian estimation of the NHPP intensity function. We have also shown an advantage of the NHPP method when we are able to assume some structure for the baseline intensity function, which may not always be possible.  These advantages of both methods should be carefully considered alongside estimation and interpretation abilities and goals.

%

Next, we evaluate the shared component model proposed in this paper with simulated data examples. We consider two cases for the distribution of the weight parameter $\delta$. In the first case, the shared component in the intensity of the first point process is weighted with $\delta$ and the shared component in the intensity of the second point process is weighted with $1-\delta$. In this case, we use a Uniform(0,1) distribution as the prior for $\delta$. In the second case, the contribution of the shared component to the intensity of the second point process is weighted by $1/\delta$, instead of $1-\delta$, and the prior on $\log(\delta)$ is a Normal distribution. The choice of the Normal distribution for the second case is motivated by the use of this distribution by \cite{knorr2001shared} in the shared component model for areal data.

From the simulated cases of the shared component model, we find that the shared component model with the Uniform prior for the weighting parameter $\delta$ provides simple interpretation and more reliable parameter estimates. We simulate from both of these weighting schemes and analyze the posterior estimates of the parameters. In Appendix Section \ref{sec:shared_comp_dist_app} we include plots of these point processes which include identical parameters and different distributions. In Table \ref{tab:shared_comp_sim_unit}, we see that the credible interval for all parameters contains the true parameter for the case when $\delta$ and $1-\delta$ are used as the weights of the shared component. In the case when $\delta$ and $1/\delta$ are used, some parameters, namely the parameter $\sigma$ (the shared component covariance function parameter) and $\beta_0$ (the intercept of the second point process) are not recovered accurately through the simulation studies. Therefore, we suggest use of the Uniform prior and the weighting scheme of $\delta$ and $1-\delta$ be used for the shared component model. 

\begin{table}
\caption{\label{tab:shared_comp_sim_unit} Estimated regression coefficients and shared component parameters with 95\% credible intervals for simulated datasets on the unit square.} \\
\centering
\adjustbox{max width=\columnwidth}{
\begin{tabular}{|c|cc|cc|cc|}
    \hline
        & \multicolumn{2}{c|}{Point Process 1 }  & \multicolumn{2}{c|}{Point Process 2} & \multicolumn{2}{c|}{Shared Component}  \\
         Model & $\beta_1$ ($x_1$) & $\beta_2$ ($x_2$) & $\beta_0$ & $\beta_1$ ($x_3$) & $\delta$ & $\sigma$ \\
         \hline
         \hline
         Truth & 0.12 & 0.06 & 0.1 & 0.25 & 0.3 & 1.7 \\
         \hline
         Estimate, Unif & 0.13  & 0.06 & 0.06  & 0.25  &  0.30 & 1.92 \\
         Credible Interval &(0.11, 0.15)& (0.05, 0.07) & (-0.07, 0.20)& (0.23, 0.28)& (0.26, 0.34) & (1.48, 2.40)\\
         \hline
         Estimate, Norm & 0.12 & 0.05 & -0.76 & 0.29 & 0.29 & 2.18 \\
         Credible Interval &  (0.10, 0.15) &  (0.04, 0.06) & (-1.21, -0.36) & (0.26, 0.31)&(0.27, 0.32) & (1.81, 2.58)\\
         \hline
\end{tabular}
}
\end{table}

\begin{figure}
    \centering
    \includegraphics[width=0.97\textwidth]{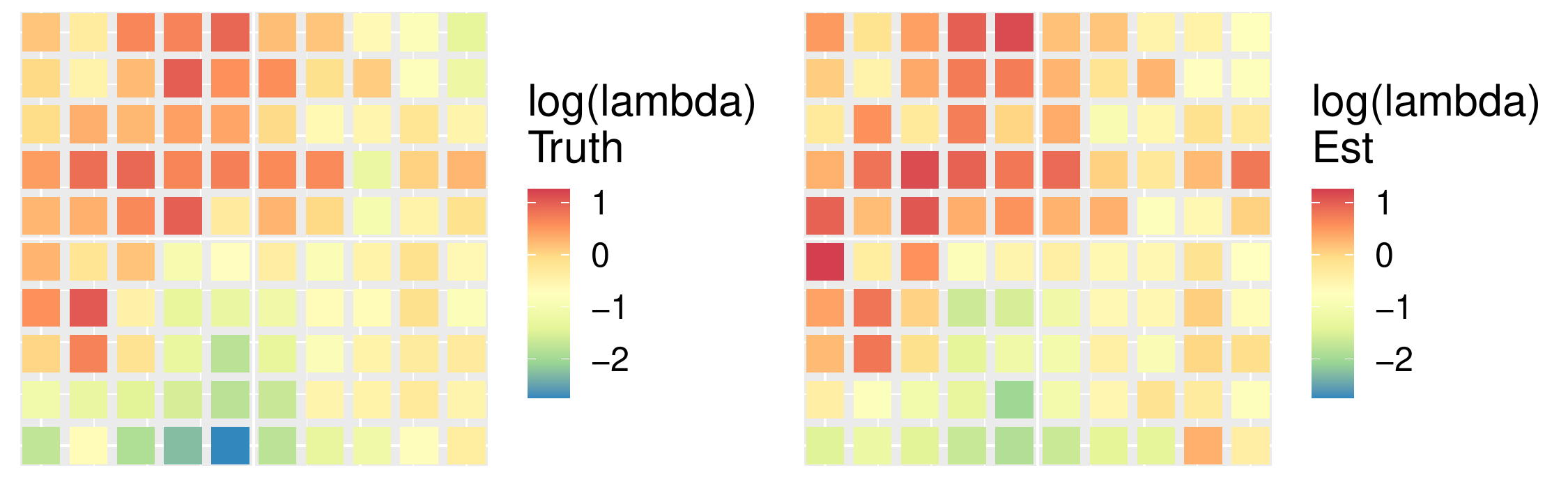}
    \caption{Shared Component: Truth and Estimated (Uniform Distribution)}
    \label{fig:shared_comp_sim}
\end{figure}

When analyzing simulated data for the shared component case, we also note that when two intercepts are used ($\beta_0$ for both point processes), there is confounding between the two parameters and they cannot be estimated accurately. Therefore, we use one intercept term for one of the point processes and omit the intercept from the other point process. We also note that we tested the inclusion of identical spatial covariates in the intensity functions for both point processes and we were able to recover accurate parameter estimates even when identical spatial covariates were included in the intensity function for both point processes, though this should be investigated in more detail. In Figure \ref{fig:shared_comp_sim}, we compare the true shared component, generated through simulation, to the posterior mean estimate of the shared component using the Uniform prior and weights of $\delta$ and $1-\delta$.  We find that we are able to recover the shared component relatively accurately.

\subsection{Results for Police Stops and Use of Force in Chicago}

Point-level datasets from Chicago allow us to utilize case-control and shared component models to create detailed analyses of the relationship between police use of force and police stops. \cite{ba2021role} aggregates the policing data we study here to a panel dataset of officer shifts and aggregate spatially by police beat. A comparison of police behavior across ``MDSBs" (month, day of week, shift, and beat) allows for comparison of officers of different demographic profiles but similar patrol assignments. \cite{ba2021role} use ordinary least squares with MSDB fixed effects to determine the effect of officer and citizen characteristics on use of force outcomes. We conduct a spatial analyses of the Chicago data used by \cite{ba2021role} but preserve the exact spatial information of both the police stops and the use of force incidents, rather than aggregating to beats. This modeling framework allows us to incorporate macroinstitutional factors, such as the decision to deploy more officers to specific neighborhoods, that has intentionally not been considered in some previous work \citep{knox2020administrative, knox2020toward}.

First, we fit the case-control model to the police use of force and stop data from Chicago through two estimation procedures: logistic regression and the spatial NHPP model, utilizing a KDE. We also estimate two different sets of spatial covariates, one with the Herfindahl Index and the unemployment rate as spatial covariates and the other substituting median age for the unemployment rate. We find that these two estimation procedures produce different results, as shown in Table \ref{tab:case_cont_chic}. The confidence interval for the median age regression coefficient includes 0 for logistic regression and the regression coefficient is estimated to be negative using the NHPP. We include the 95\% confidence intervals for logistic regression and the 95\% credible intervals for the NHPP. We note that the credible intervals do not overlap in the majority of cases. We also evaluate model fit for the NHPP using WAIC and using AIC for logistic regression, where lower WAIC and AIC both indicate better model fit. We find that the model fit checks indicate the opposite model has a better fit between the two estimation procedures. Many open questions remain from this analysis of the use of force and stop data from Chicago using two estimation procedures for the same model. We find that it is perhaps safest to interpret the regression coefficients obtained from fitting a logistic regression here in terms of the factors influencing an event being a case versus being a control. This is in contrast to interpreting the results as factors influencing the spatial distribution of cases, scaled by controls. %

\begin{table}
\caption{\label{tab:case_cont_chic} Estimated regression coefficients with 95\% confidence/credible intervals (logistic regression/ NHPP, respectively) for the case-control model applied to the Chicago police use of force and police stop data. } \\
\centering
\adjustbox{max width=1\columnwidth}{
\begin{tabular}{|c|ccc|c|}
    \hline
         Model & $\beta_0$ & $\beta_1$ (Herf index) & $\beta_2$ (unemp rate) & WAIC/AIC \\
         \hline
         NHPP & -2.10 (-2.17, -2.03) & -0.88 (-1.01, -0.76) & -0.49 (-0.74, -0.24) & 145,244 \\
         Log Reg & -4.89 (-4.96, -4.82) &  -0.62 (-0.74, -0.50) &  -1.19 (-1.45, -0.93) & 94,338 \\
         \hline
         Model & $\beta_0$ & $\beta_1$ (Herf index) & $\beta_2$ (med age) & WAIC/AIC \\
         \hline
         NHPP & -1.75 (-1.89, -1.62) & -0.83 (-0.94, -0.72) & -0.01 (-0.02, -0.01) & 145,210 \\
         Log Reg & -5.21 (-5.35, -5.07) &  -0.34 (-0.44, -0.23) &  0.001 (-0.003,  0.005) & 94,420 \\
         \hline
\end{tabular}
}
\end{table}

Next, we analyze the Chicago police stop and use of force datasets using the shared component model. We adopt the approach of \cite{xu2019stochastic} in sampling one of the point processes. We spatially thin the stop dataset due to the large number of points in this dataset, where each point has an equal probability of remaining in the dataset. In our case, we consider the thinning probability to be 10\%. %
In the Appendix, Section \ref{sec:chic_sample_stops_app}, we describe this approach in detail. In our analysis, we are only interested in describing the spatial distribution and the spatial relationship between use of force incidents and police stops. If we were interested in any other characteristics of the stop data, such as the demographics of the citizens or officers involved, we would want to consider either using the full stop data or a different sampling approach that takes into account these variables of interest.

In Table \ref{tab:shared_comp_chic}, we include the estimated coefficients from the intensity functions for both point processes and the parameters associated with the shared component, $\sigma$ and $\delta$. We apply both parameterizations of the weight term, $\delta$, described in Table \ref{tab:shared_comp_chic} by the prior distribution (either Normal or Uniform).  We see that the estimates between the two different parameterizations of the shared component model are similar for the covariate effects, $\bm{\beta}$. Both the point estimates and the credible intervals are very similar across parameterizations. We note that the $\delta$ parameter for both parameterizations indicates that the shared component contributes most strongly to the use of force dataset, rather than the police stop data. 
As in \cite{knorr2001shared}, we assume that the shared component model is indicative of spatial variation in factors influencing both police use of force and stops, such as increased police activity or population. Therefore, we interpret $\delta$ for both parameterizations as showing that the shared spatial pattern, due to factors such as police activity or population, affects the distribution of police use of force events more so than police stops.
The only parameter that notably differs between the two parameterizations is $\sigma$, from the covariance function of the Gaussian process, which is estimated to be higher in the case when $\delta$ is bounded between 0 and 1 with a Uniform prior.

\begin{table}
\caption{\label{tab:shared_comp_chic}  Estimated regression coefficients with 95\% credible intervals for the shared component model applied to the Chicago police use of force and police stop data. } \\
\centering
\adjustbox{max width=\columnwidth}{
\begin{tabular}{|c|cc|cc|cc|}
    \hline
        & \multicolumn{2}{c|}{Point Process 1 - UoF}  & \multicolumn{2}{c|}{Point Process 2 - Stops} & \multicolumn{2}{c|}{Shared Component}  \\
         Model & $\beta_1$ (Herf index) & $\beta_2$ (unemp rate) & $\beta_0$ & $\beta_1$ (med age) & $\delta$ & $\sigma$ \\
         \hline
         Estimate, Unif & -0.52 &-1.4 & 4.44 & -0.017 & 0.54& 6.31  \\
         Credible Interval & (-0.64, -0.39) & (-1.7, -1.07) &(4.24, 4.67) & (-0.018, -0.016) &  (0.53, 0.55) & (4.91, 7.87) \\
         
         \hline
         Estimate, Norm & -0.51 & -1.38 & 4.44& -0.017& 1.09 &3.14 \\
         Credible Interval & (-0.65, -0.37)& (-1.67, -1.06)& (4.15, 4.66) & (-0.019, -0.016) &  (1.07, 1.11) & (2.46, 3.92)\\

         \hline
\end{tabular}
}
\end{table}

\begin{figure}
    \centering
    \includegraphics[height=0.28\textwidth]{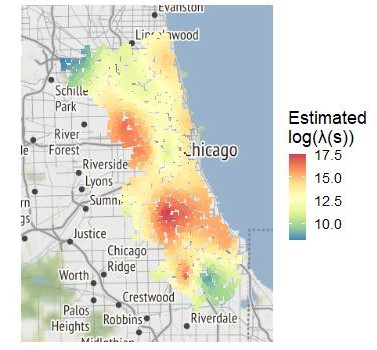}
    \includegraphics[height=0.28\textwidth]{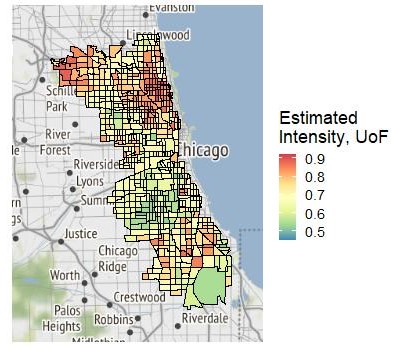}
    \includegraphics[height=0.28\textwidth]{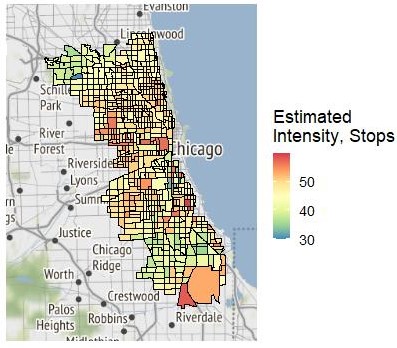}
    \caption{Shared Component (left) and Point Process-Specific Components (use of force- middle, police stops- right), Based on Posterior Mean Estimates}
    \label{fig:shared_comp_chic}
\end{figure}

We also would like to compare the estimated spatial shared component to the point process-specific components. In Figure \ref{fig:shared_comp_chic} we plot the inferred spatial distribution for all facets of the shared component model. We note that the estimated spatial distribution for the point process-specific part of the model in this case is a nonhomogeneous Poisson process, so the estimated intensity only varies with the census tracts. This is illustrated in the middle and right plots of Figure \ref{fig:shared_comp_chic}. In future work, one could consider incorporating an LGCP into this framework, although there may be problems due to confounding between the spatially varying shared component and the Gaussian process unique to each point process. In Figure \ref{fig:shared_comp_chic}, we plot the posterior mean estimate for the shared component on the left side, transformed over the integration points used in our analysis. We see that there is a notable shared spatial pattern between the two point processes. We note that we have not used population as a spatial covariate in our model. Therefore this shared spatial component may largely be determined by population and/or increased police activity in these areas, as noted earlier. In the future, we could consider also scaling both intensity functions by population density, as is done frequently for point process intensity functions \citep[cf.][]{liang2008analysis, walder2020privacy}. 
Next, we turn our attention to the spatial components unique to each point process, which indicate the existence of factors that are relevant to only police of force or police stops, but not both. We analyze the posterior mean estimates for the regression coefficients and we note that the distribution of police use of force incidents beyond the shared component is most notable on the northern side of the city, with some higher intensity spots in the southern side of the city as well. On the other hand, the police stop intensity is highest in the far south and middle parts of the city, after accounting for the shared component. This information can inform our analysis of where use of force and police stops are higher, after accounting for the other shared spatial pattern. Future research could help determine additional spatial variables and socioeconomic information that could be included in the model that would inform the factors that influence one point process, but not both. This could help identify why police stops occur more frequently in some areas further than what we would expect from police activity alone, and could answer the same question for police use of force. %

\section{Discussion}
\label{section_discussion}

We have demonstrated through Chicago policing data and simulated examples that the shared component model developed here can provide new insights when relating two point processes. In practice a causal direction between the two processes is often not justified. Our model provides a mechanism to compare two point processes without specifying which process is a case or control, thereby avoiding the requirement that modelers impose a causal direction. 
We can fully characterize spatial patterns common to both point processes as well as those unique to each point pattern. The application to Chicago policing data provides a particularly useful context in which to tease apart spatial patterns that are common to use of force and police stops as well as patterns that are unique to each point process. Our approach involves the use of a Gaussian process, which is more computationally complex than the case-control model, but provides additional flexibility. We also suggest methods to decrease this computational burden, such as sampling one of the point process, as suggested by \cite{xu2019stochastic}.

The shared component model allows us to create a rich characterization of the relationship between two point processes. Instead of scaling one point process intensity by the intensity of another point process, we can incorporate spatial/community variables when determining the possible unique effects on these point processes, after determining drivers for both point processes, such as population. This allows us to analyze potential drivers of both point processes, rather than just one of the point processes. This model also allows for visualization of the spatial trends of both point processes, as well as drivers of both point processes that have been inferred from the data.

In this work, we focus our analysis on the comparison of the spatial distribution between two point patterns: police use of force and police stops. This framework could be expanded in the future to more than two point processes by allowing the weight parameter of the shared component, $\delta$, to adapt to more than two point processes. This analysis does not attempt to analyze the relationship between officer and citizen characteristics and the spatial patterns of police behavior, though this could be achieved by combining this method with existing methods, such as those developed by \cite{kelling2021two}. We also note that the shared component could take forms other than a Gaussian process, for instance by using a clustering model \citep{knorr2001shared}. %

\section{Acknowledgements}
The authors would like to thank Professor Peter Diggle and Professor Ephraim Hanks for helpful conversations that greatly improved the manuscript. This project was supported by Award No. 2020-R2-CX-0033, awarded by the National Institute of Justice, Office of Justice Programs, U.S. Department of Justice. The opinions, findings, and conclusions or recommendations expressed in this publication are those of the authors and do not necessarily reflect those of the Department of Justice.

\section{Appendix}

\subsection{Integration Points}
\label{sec:integ_methods_app}

\begin{figure}
    \centering
    \includegraphics[width=0.52\textwidth]{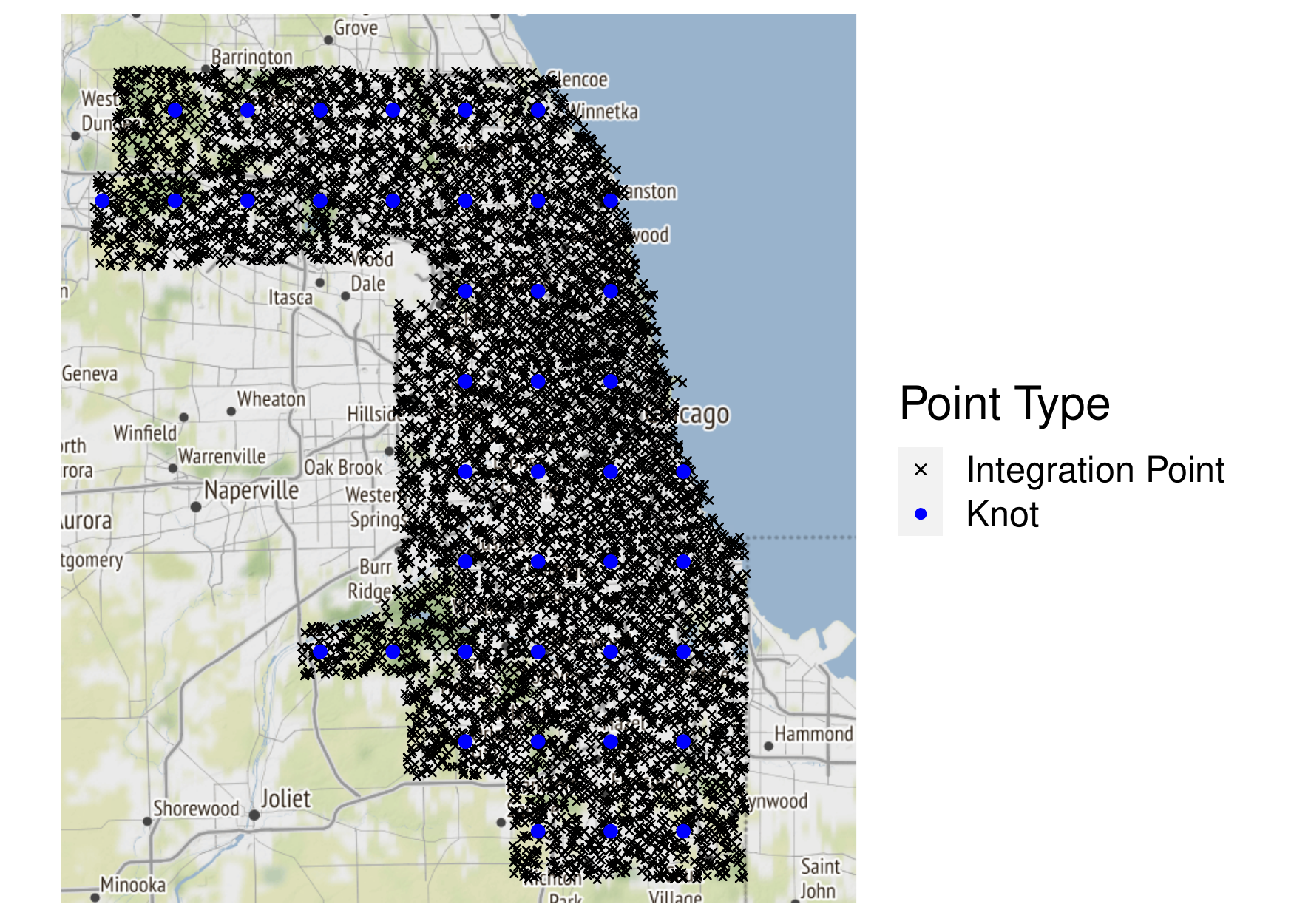}
    \caption{Integration Points, Predictive Process Points}
    \label{fig:integ_predproc_points}
\end{figure}

Unlike other urban areas, Chicago has many very small census tracts. In order to accurately integrate the intensity function for Chicago, we must include at least one point per census tract and the number of integration points per census tract must be proportional to the area of the census tract, so that all census tracts are weighted by their area when integrating the intensity function over the region \citep[cf.][]{liang2008analysis}. We use a total of 7,304 integration points for Chicago. Alternative parameterizations, such as the number of points being proportional to the square root of the area, create inaccurate estimates of the integral for Chicago, due to the numerous small census tracts. We have shown the integration points as well as the knots used for the predictive process for the spatially varying shared component in Figure \ref{fig:integ_predproc_points}. 

This approach of having the number of integration points per census tract being exactly proportional to the area is particularly important for Chicago, where the difference in area between the smallest census tracts is quite extreme. For an example, we compare Dallas, Texas and Chicago, Illinois, both of which contain a range of sizes for census tracts. The largest census tract in the Dallas city limits is 115 times bigger than the smallest census tract. For Chicago, the largest tract is 3,573 times bigger than the smallest tract. We also analyze more typical values, rather than the extremes, through the first and third quartiles of the census tract areas. The third quartile of the Dallas census tracts is 2.8 times bigger than the first quartile while the third quartile of the Chicago census tracts is 5.3 times bigger than the first quartile.

\subsection{Shared Component Weighting Distribution}
\label{sec:shared_comp_dist_app}

\begin{figure}
    \centering
    \includegraphics[width=0.6\textwidth]{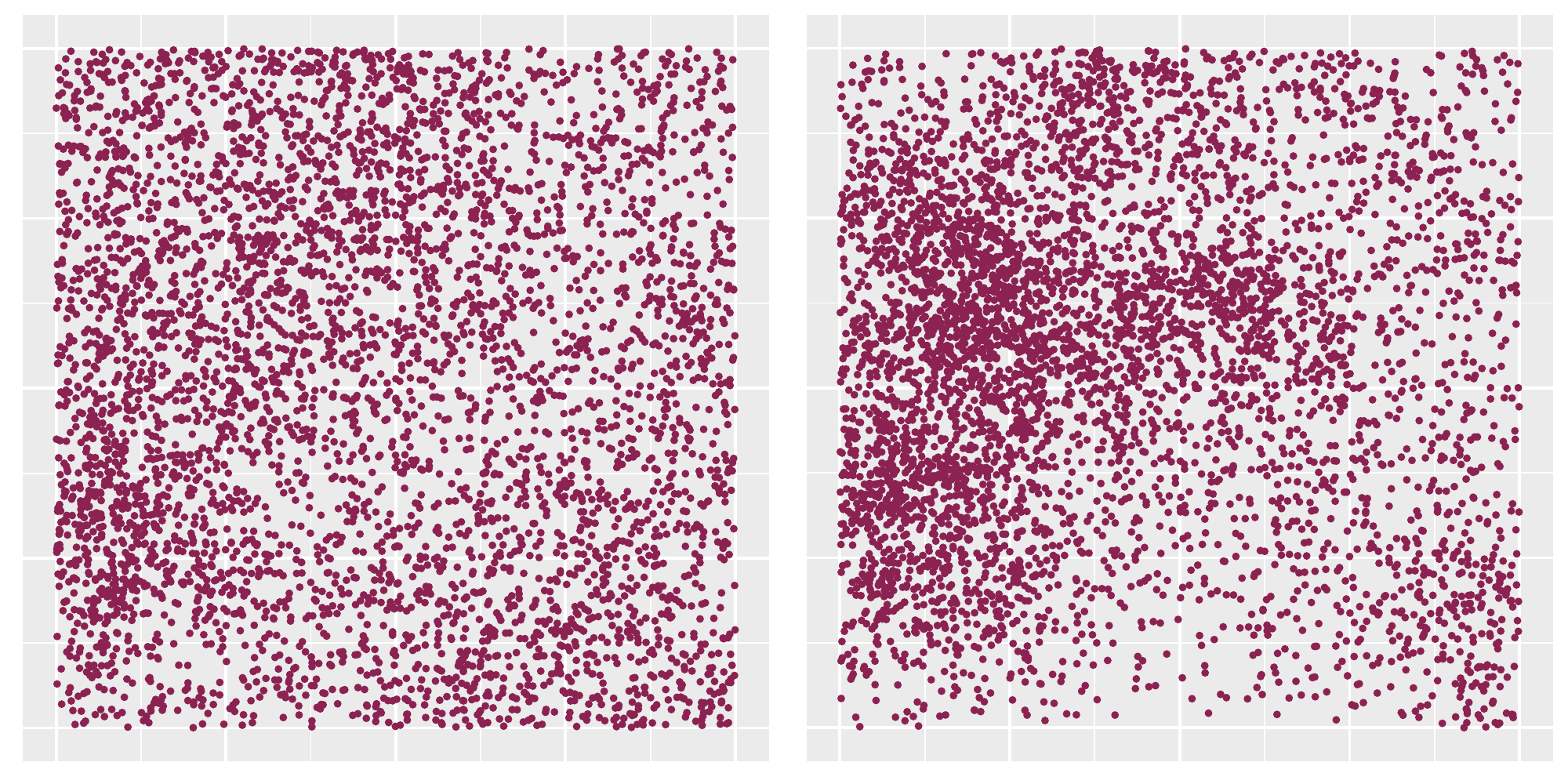}
    \includegraphics[width=0.6\textwidth]{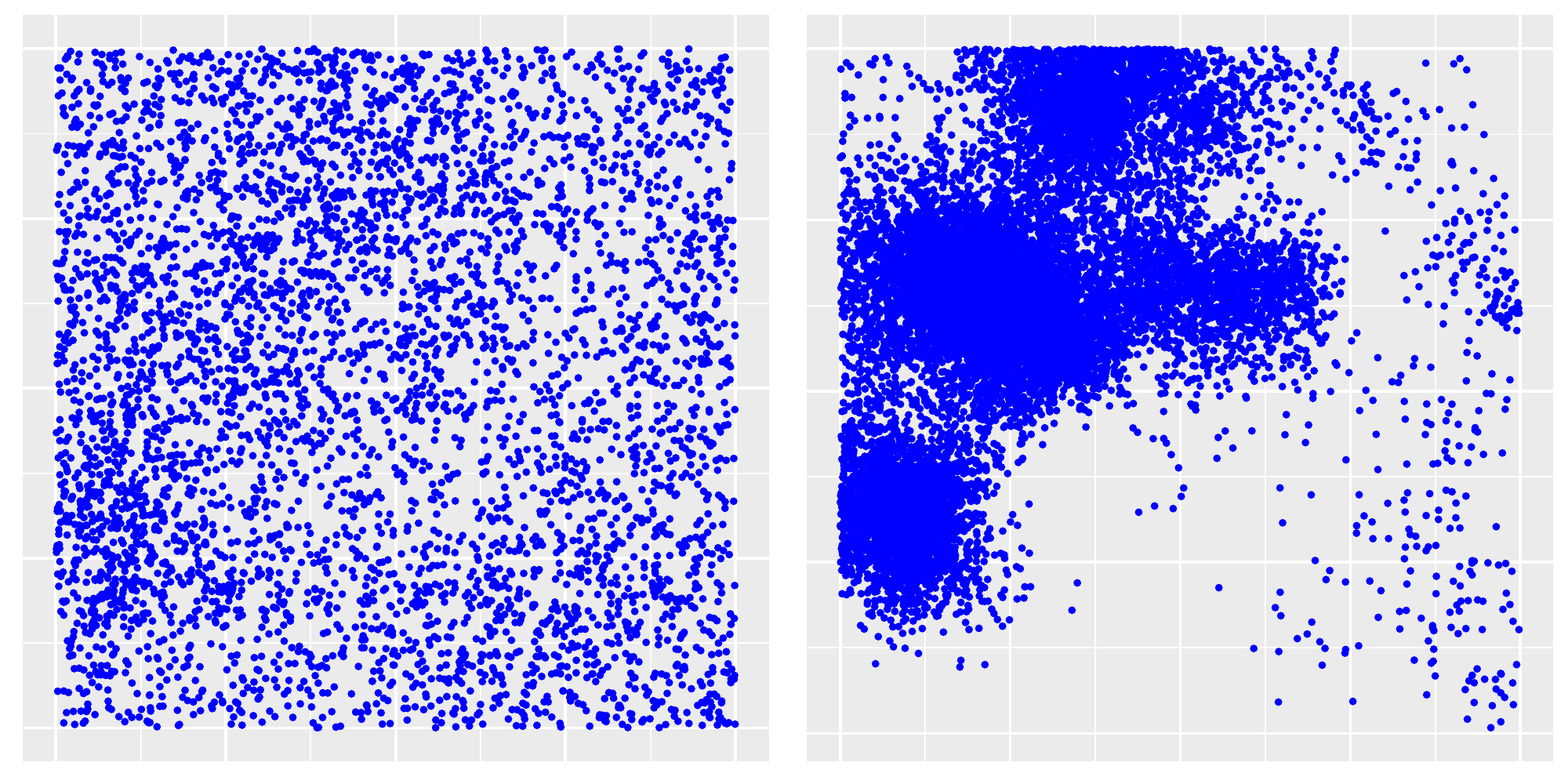}
    \caption{Comparison of uniform (top) and normal (bottom) distributions for shared component, all other parameters equal}
    \label{fig:shared_comp_dist_comp}
\end{figure}

We simulate two point processes using identical parameters, as shown in Table \ref{tab:shared_comp_sim_unit}, but different weighting schemes and distributions. These are based on spatial covariates shown in Figure \ref{fig:sim_spat_var}. In magenta in Figure \ref{fig:shared_comp_dist_comp}, we show the point process that results from the weight contributions $\delta$ and $1 - \delta$ to the two point processes, with a Uniform(0,1) prior for the $\delta$ parameter. In blue, on the bottom of Figure \ref{fig:shared_comp_dist_comp}, we show the two point processes that result from the weights $\delta$ and $1/\delta$ with a Normal prior for the $\delta$ parameter. In both simulations, $\delta$ was set to be 0.3.

\subsection{Sample of Chicago Stops}
\label{sec:chic_sample_stops_app}

\begin{figure}
    \centering
    \includegraphics[height=2in]{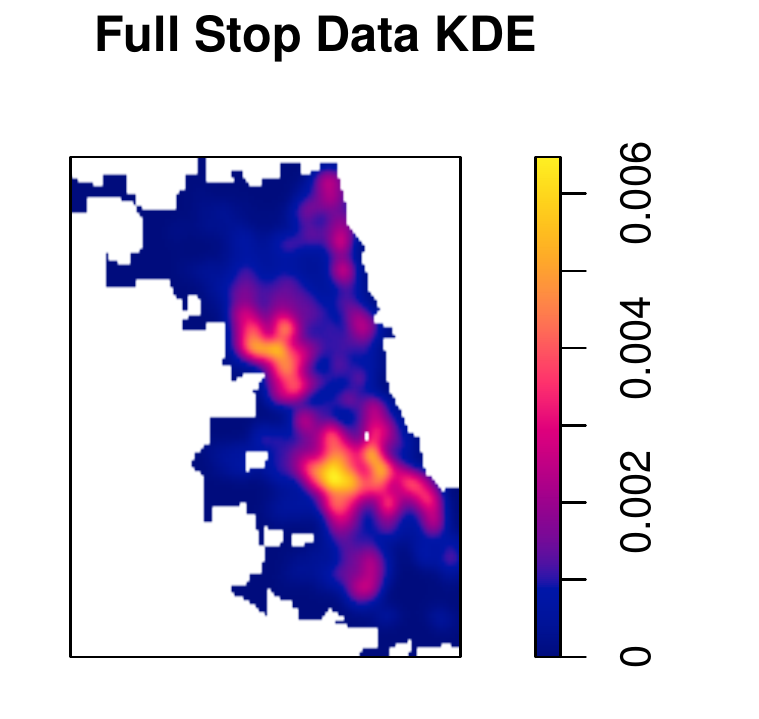}
    \includegraphics[height=2in]{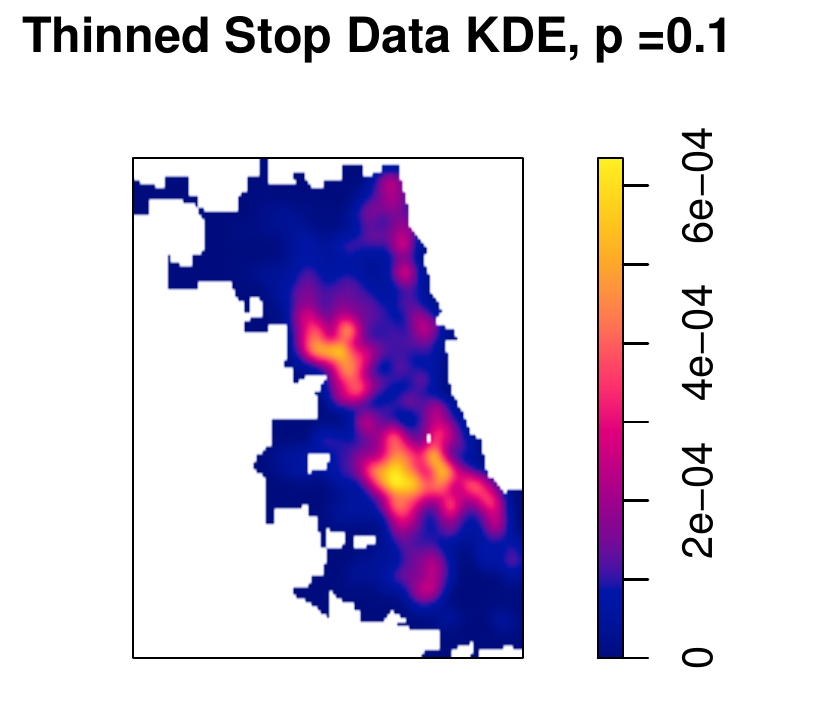}
    \caption{KDE of full stop data (left) compared to thinned stop data (right)}
    \label{fig:stop_kde_comp}
\end{figure}

For the shared component model, we must transform the shared component (involving a Gaussian process) to be the dimension of both the stop data and the use of force data. The police use of force data has 7,539 events with complete spatial information, which is manageable computationally. The stop data, on the other hand, contains 1,453,832 events with complete spatial information. This represents a computational challenge in terms of large matrix computations. We notice that if we perform uniform thinning across all points, the spatial distribution of points does not change much between the full stop data and the thinned data. In Figure \ref{fig:stop_kde_comp}, we show the kernel density estimate (KDE) for the full stop data next to the thinned stop data, where the probability of keeping each point is 10\%. The spatial thinning results in a dataset of 145,783 points. We notice that the KDE looks almost identical between the two point processes, except for the scale, shown to the right of the plot. We adopt a similar approach to \cite{xu2019stochastic} where we adopt sampling of one of the point processes. In future work, when we are interested in analyzing different features of the stops, we may want to consider the full point process or a more complex sampling mechanism that samples based on other variables as well.

\subsection{Relationship Between Case-Control and Shared Component Model}
\label{sec:app_case_cont_shared}

In this section, we elaborate on the relationship between the case-control model \citep[cf.][]{diggle2007second} and our shared component model for point processes. The case-control model defines the intensity of a case process, $\lambda_1^{CC}(s)$, based on the intensity of a control process,  $\lambda_0^{CC}(s)$, as follows %
\begin{equation}
    \begin{array}{l}
    \lambda_1^{CC}(s) = \lambda_0^{CC}(s)\exp(\bm{z(s)}'\bm{\beta}). 
    \end{array}
    \label{case_cont_app}
\end{equation}

As a reminder the shared component model, shown below, %
describes the intensities of two point processes separately (denoted $\lambda_1(s)$ and $\lambda_2(s)$) but relies on a process that contributes to the intensity of both processes, which we have denoted $\lambda(s)$. 
\begin{equation}
    \begin{array}{l}
    \lambda_1(s) = \lambda(s)^\delta \exp\left(\bm{z_1(s)}'\bm{\beta_1} \right) \\ \newline
     \lambda_2(s) = \lambda(s)^{1-\delta} \exp\left(\bm{z_2(s)}'\bm{\beta_2}\right)  
    \end{array}
    \label{shared_comp_eq_app}
\end{equation}
This structure of a shared process contributing to both intensities allows us to write the shared component model as a version of the case-control model. 

Our goal is to determine the parameterization of the `scaling process' for the shared component model, which is simply the control process  $\lambda^{CC}_0(s)$ for the typical case-control model.
In Equation \ref{shared_comp_case_cont_app}, we rewrite the intensity for the first point process in the shared component model to show the scaling process under the structure of the case-control model. We do this by solving for the shared component that contributes to both point processes, $\lambda(s)$, in the intensity of the second point process and substituting this in the intensity of the first point process. We denote spatial covariates as $\bm{z_1(s)}$ and their corresponding regression coefficients as $\bm{\beta_1}$ for the first point process and spatial covariates as $\bm{z_2(s)}$ and regression coefficients as $\bm{\beta_2}$ for the second point process. Instead of purely being an estimate of the second spatial intensity function as in the case-control model ($\lambda_0^{CC}(s)$), now this scaling process for the re-parameterized shared component model $\left(\left[\frac{\lambda_2(s)}{\exp(\bm{z_2(s)}'\bm{\beta_2})}\right]^{\delta/(1-\delta)}\right)$ depends on the second spatial intensity, the covariates and coefficients for the second spatial intensity, and the parameter $\delta$ which defines the contribution of the shared component to both point processes.

\begin{equation}
\lambda_1(s) = \left[\frac{\lambda_2(s)}{\exp(\bm{z_2(s)}'\bm{\beta_2})}\right]^{\delta/(1-\delta)} \exp(\bm{z_1(s)}'\bm{\beta_1}). %
    \label{shared_comp_case_cont_app}
\end{equation}

\bibliographystyle{rss}
\bibliography{references}
\end{document}